\documentclass[preprint2, tighten]{aastex63}

\usepackage{graphicx}
\usepackage{latexsym}
\usepackage{amsfonts,amsmath,amssymb}
\usepackage{epstopdf}
\usepackage{natbib}
\usepackage{mathrsfs}
\usepackage{algorithm}
\usepackage[utf8]{inputenc}
\usepackage{hyperref}
\usepackage{xspace}
\usepackage{bibentry}
\usepackage{diagbox}

\received{}
\revised{}
\accepted{}
\submitjournal{ApJL}

\shorttitle{MC Reconnection}
\shortauthors{Gaches et al.}

\begin{document}

\defcitealias{lv1999}{LV99}

\title{CRAFT (Cosmic Ray Acceleration From Turbulence) in Molecular Clouds}

\correspondingauthor{Brandt A. L. Gaches}
\email{gaches@ph1.uni-koeln.de}

\author[0000-0003-4224-6829]{Brandt A. L. Gaches}
\affiliation{I. Physikalisches Institut, Universit\"{a}t zu K\"{o}ln, Z\"{u}lpicher Stra{\ss}e 77, 50937, K\"{o}ln, Germany}
\affiliation{Center of Planetary Systems Habitability, The University of Texas at Austin,  USA}

\author[0000-0001-6941-7638]{Stefanie Walch}
\affiliation{I. Physikalisches Institut, Universit\"{a}t zu K\"{o}ln, Z\"{u}lpicher Stra{\ss}e 77, 50937, K\"{o}ln, Germany}
\affiliation{Center for Data and Simulation Science, Universit\"{a}t zu K\"{o}ln, K\"{o}ln, Germany}

\author{A. Lazarian}
\affiliation{Department of Astronomy, University of Wisconsin-Madison, USA}

\begin{abstract}
Low-energy cosmic rays, in particular protons with energies below 1 GeV, are significant drivers of the thermochemistry of molecular clouds. However, these cosmic rays are also greatly impacted by energy losses and magnetic field transport effects in molecular gas. Explaining cosmic ray ionization rates of $10^{-16}$ s$^{-1}$ or greater in dense gas requires either a high external cosmic ray flux, or local sources of MeV-GeV cosmic ray protons. We present a new local source of low-energy cosmic rays in molecular clouds: first order Fermi-acceleration of protons in regions undergoing turbulent reconnection in molecular clouds. We show from energetic-based arguments there is sufficient energy within the magneto-hydrodynamic turbulent cascade to produce ionization rates compatible with inferred ionization rates in molecular clouds. {\it As turbulent reconnection is a volume-filling process, the proposed mechanism can produce a near-homogeneous distribution of low-energy cosmic rays within molecular clouds}.

\end{abstract}


\section{Introduction}
Molecular clouds are immersed in a bath of cosmic rays (CRs), i.e. energetic charged particles that are accelerating and propagating in our galaxy \citep{schlickeiser2002}. Low-energy particles, particularly protons with energies between 1 MeV to 1 GeV, influence the thermochemistry of molecular gas in regions which are well-shielded from ultraviolet radiation \citep[see reviews by][]{dalgarno2006, padovani2020}. The chemistry of cold molecular gas is regulated through ion-neutral reactions. Ion-neutral chemistry is largely initiated following the ionization of H$_2$ and subsequent production of H$_3^+$ 
\begin{equation}
    {\rm H_2 + H_2^+ \rightarrow H_3^+ + H.}
\end{equation}
Deuterium chemistry is also regulated in a similar manner through the production of H$_2$D$^+$ from HD. The CR ionization rate (CRIR), $\zeta$, is inferred primarily through molecular line features from ions, such as H$_3^+$, OH$^+$ and H$_{\rm n}$O$^+$ absorption and emission from species such as HCO$^+$, DCO$^+$ and N$_2$H$^+$. These observations have shown that the CRIR in diffuse molecular gas spans $10^{-16} < \zeta < 10^{-15}$ s$^{-1}$ \citep[e.g.][]{indriolo2012, indriolo2015, neufeld2017}. In the dense gas, observations infer the CRIR in a wider range from $10^{-17} < \zeta < 10^{-15}$ s$^{-1}$ \citep[e.g.][]{caselli1998, favre2017, barger2020}. Recently, observations and astrochemical models have inferred an CRIR towards the protocluster OMC-2 FIR 4 of approximately $\zeta \approx 10^{-14}$ s$^{-1}$ \citep{ceccarelli2014, favre2018}. 

One-dimensional models of transport through molecular gas conflict with the heightened ionization rates inferred in shielded, dense gas. These models ubiquitously predict a declining CRIR with column density, $\zeta(N)$, whether due to energy-losses\footnote{The dominant energy losses for 1 MeV - 1 GeV protons are due to ionizing atomic and molecular material and the production of pions.} or magnetic-field effects\footnote{Magnetic-field effects such as screening, mirroring and streaming instabilities.} \citep{padovani2009, morlino2015, schlickeiser2016, ivlev2018, phan2018, silsbee2019, fujita2021}. A dense gas CRIR similar to that inferred in diffuse gas may necessitate a source of low-energy CRs within the gas. Some of these sources have already been posited, such as protostellar jets \citep{padovani2016}, protostellar accretion shocks \citep{padovani2016, gaches2018}, HII regions \citep{meng2019, padovani2019} and embedded stellar winds \citep{yang2020}. These are localized sources of CRs acceleration that are expected to induce rather inhomogeneous distribution of CRs within molecular clouds.

We propose a new source of low-energy CRs in dense magnetized molecular clouds: particles accelerated within zones of turbulent reconnection. Turbulence is known to be part and parcel of the molecular cloud dynamics \citep{mckee2007} and it is known to be accompanied by fast turbulent reconnection \citep{lv1999, eyink2011, eyink2013}. The latter process is known to induce the acceleration of energetic particles \citep{pino2005, kowal2011, lazarian2020}.   

We explain in Section \ref{sec:model} our simplified model of CR production in reconnection zones in magneto-hydrodynamic turbulence. Section \ref{sec:results} presents the results of these calculations and discusses the broad implications of the mechanism.

\section{Method}\label{sec:model}
\begin{figure*}[ht!]
    \centering
    \includegraphics[width=0.75\textwidth]{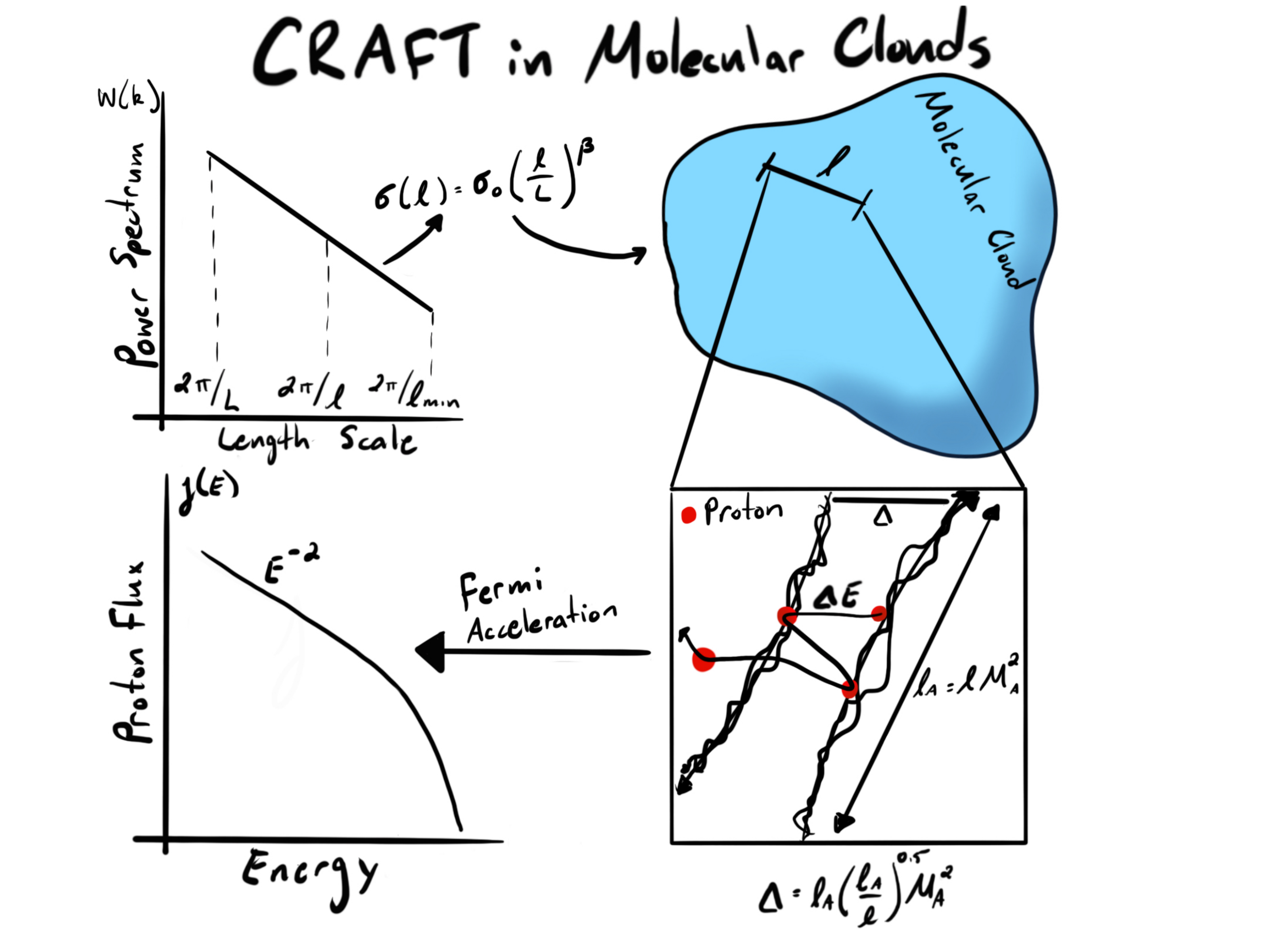}
    \caption{\label{fig:schematic} Basic schematic of the proposed mechanism. Top left: We assume a power-law power spectrum related to a line width-size relation. Top right: The turbulence driven at scale, $\ell$, corresponds to an average density and magnetic field ``seen'' by the turbulence. Bottom right: Within the turbulence there are reconnection regions of width $\Delta$ (Eq. \ref{eq:delta}). Here, protons bounce between the reconnection fronts. Bottom left: The resulting acceleration, via the Fermi mechanism, results in a power-law energy distribution. }
\end{figure*}

We assume a spherical, magnetized, hierarchical cloud with magneto-hydrodynamic (MHD) turbulence. MHD turbulence is injected at some large scale, $L$, with a velocity dispersion, $\sigma_0$, and cascades through the cloud. Figure \ref{fig:schematic} gives a schematic of the proposed mechanism. At a given length scale, $\ell$, the turbulent linewidth is given by the linewidth-size relation \citep{larson1981, mckee2007, heyer2015}
\begin{equation}
    \sigma(\ell) \approx \sigma_0 \left ( \frac{\ell}{L}\right )^{\beta}
\end{equation}
where we use $\beta = 0.5$. The density of the gas is calculated by assuming the gas can be  prescribed by the virial parameter, $\alpha_V$, defined by
\begin{equation}
    \alpha_V = \frac{5\sigma^2 \ell}{G M}
\end{equation}
where $G$ is the gravitational constant. The density is thus
\begin{equation}
    \rho = \frac{15}{4\pi G\alpha_V} \left ( \frac{\sigma^2}{\ell^2}\right ).
\end{equation}
Finally, the magnetic field is calculated using the empirical fit from \citet{crutcher2019}
\begin{equation}
    B(n) = 
    \begin{cases}
    B_0 & n \le n_0 \\
    B_0 \times \left ( \frac{n}{n_0}\right)^{\kappa}& n > n_0
    \end{cases}
\end{equation}
where $n = \rho/(2.33 \times m_H)$ is the number density, $B_0$ = 10 $\mu$G, $n_0$ = 300 cm$^{-3}$ and $\kappa$ = 0.65. 

The energy within the MHD turbulence cascade depends on the Alfv\'{e}n Mach number, $\mathcal{M}_A = \sigma(\ell)/v_A$ where $v_A$ is the Alfv\'{e}n speed, $v_a = \frac{B}{\sqrt{4\pi\rho}}$. In the ideal-MHD case, the dissipiation rate of the specific energy per unit mass is given by \citep{lazarian2020}
\begin{equation}
    \epsilon = 
    \begin{cases}
        \frac{\sigma^3}{\ell} & \mathcal{M}_A \ge 1 \\
        \frac{\sigma^4}{v_A \ell} & \mathcal{M}_A < 1.
    \end{cases}
\end{equation}
The dissipation rate of energy per unit volume is then $\varepsilon = \rho \epsilon$. A fraction of this energy, $f_{\rm CR}$ goes into CR acceleration, such that
\begin{equation}
    \varepsilon_{\rm CR} = f_{\rm CR} \varepsilon,
\end{equation}
where we take $f_{\rm CR} = 0.01$ as an estimated lower limit of the acceleration efficiency. We assume particles are accelerated within the turbulent reconnection regions via a first-order Fermi process \citep{dgdp2005}. Following \citet{khiali2015}, the CRs are isotropically injected with an exponentially suppressed power law
\begin{equation}
    Q(E) = Q_0 E^{-\gamma} e^{-\frac{E}{E_0}}
\end{equation}
where we take $\gamma = 2$ and $E_0 = 10$ GeV\footnote{This limit was determined by examining the energy-loss and acceleration timescales}. Increasing $E_0$ negligibly impacts our main results, due to the weak dependence of the CRIR on super-GeV CRs. Further, changing $\gamma$ between $2$ and $3/2$ produces no qualitative changes in the results, nor quantitative variations over an order of magnitude. The normalization factor, $Q_0$, is calculated by assuming
\begin{equation}
    \varepsilon_{\rm CR} = \int dV \int_{E_{\rm min}}^{E_{\rm max}} Q(E) dE
\end{equation}
where E$_{\rm min} = 13.6$ eV and $E_{\rm max} = 100$ GeV. These bounds have a minor impact on the overall results of the work. Determining the injection and maximum energies requires particle-in-cell calculations of the CR acceleration and injection within molecular cloud reconnection zones. However, even if energy losses are ignored, the necessary acceleration timescale from $E_{\rm max} = e B v_A^2 \delta t$ to accelerate protons up to 100 GeV exceeds molecular cloud lifetimes for much of the parameter space.

The CR proton spectrum from the reconnection zones is a balance of injection and energy-losses. The steady-state energy-loss solution \citep{longair2011} for the number density of protons within the reconnection region, $\mathcal{N}_p(E)$, is
\begin{equation}
    \mathcal{N}_p(E) = \left | \frac{dE}{dt} \right |^{-1} \int_E^{E_{\rm max}} Q(E) dE
\end{equation}
where $\frac{dE}{dt}$ is calculated using a prescribed loss function, $\mathcal{L}(E)$
\begin{equation}
    \frac{dE}{dt} = \mathcal{M}_s^2 n v_{\rm CR}(E) \mathcal{L}(E)
\end{equation}
and $\mathcal{M}_s$ is the sonic Mach number, $\mathcal{M}_s = \sigma(l)/c_s$, $c_s = \sqrt{k_b T/\mu m_H}$, $\mu$ is the mean molecular weight and $T = 10$ K, and $v_{\rm CR}(E)$ is the relativistic velocity of the CR. We utilize the loss function given in \citet{padovani2009}. 

Turbulent reconnection is an essential part of the turbulent cascade (Lazarian \& Vishniac 1999) and a volume-filling process. This induces in CR acceleration and we model the resulting CR number density accelerated by turbulent reconnection at length scale $\ell$ by assuming the CRs diffuse from the reconnection zones and undergo energy losses. We assume an energy-dependent empirical diffusion coefficient\footnote{The process of CRs diffusion in MHD turbulence is pretty complicated with different components of MHD modes acting very differently on CRs \citep[see][]{yan2004} and its effects for the diffusion parallel and perpendicular to the mean magnetic field is also not trivial \citep[see][]{lazarian2014}. However, for the same of simplicity, in this paper we adopt the simplest possible assumptions about the diffusion. This assumption is further justified by \citet{lazarian2021} which presented a new non-resonant scattering process in turbulent magnetized media.} \citep{longair2011}:
\begin{equation}
D(E) = D_0 \left ( \frac{E}{10 \, {\rm GeV}} \right )^{\delta} {\rm \, cm^2 s^{-1}}
\end{equation}
using $\delta = 0.5$ and different values of $D_0$. The diffusion length scale is defined by
\begin{equation}
\ell_D = D(E)/v_{\rm CR}(E).
\end{equation}
The energy-loss scale, or the range, $R(E)$ is given by the stopping column, \citep{padovani2009}
\begin{equation}
n\times R(E) = \int_0^E \frac{dE}{\mathcal{L}(E)} \, \rm{cm^{-2}}.
\end{equation}
We then define a transport length scale, $\ell_T$,
\begin{equation}
    \ell_T^{-1} = \ell_D^{-1} + R^{-1}.
\end{equation}
Finally, we use the volume-filling fraction of the reconnection zones
\begin{equation}
    f_V = V_{\rm rec}/V_\ell
\end{equation}
where $V_{\rm rec}$ is the volume of a sheet-like reconnection zone, 
\begin{equation}
    V_{\rm rec} = l_A^2\Delta,
\end{equation}
$l_A = \ell M_A^{-3}$ for super- and trans-Alfv\'{e}nic turbulence and $l_A = \ell M_A^2$ for sub-Alfv\'{e}nic turbulence \citep{lazarian2020}. Following \citet{lv1999} the reconnection zone width is
\begin{equation}\label{eq:delta}
\Delta = l_A \left ( \frac{l_A}{\ell}\right )^{0.5} \mathcal{M}_A^2
\end{equation}
The volume of a region of radius, $\ell$ is $V_l = \frac{4}{3}\pi \ell^3$. The number density of transported CRs is
\begin{equation}
    \mathcal{N}_T(E) = \mathcal{N}_p e^{-\frac{\ell}{\ell_T}}.
\end{equation}

Either number density, $\mathcal{N}_{p}(E)$ or $\mathcal{N}_{T}(E)$, can be converted to a flux by
\begin{equation}
j_{\{p, T\}}(E) = \frac{v_{\rm CR}(E)\mathcal{N}_{\{p,T\}}(E)}{4\pi}.
\end{equation}
The resulting CRIR due to turbulent reconnection at length scale, $\ell$, is 
\begin{equation}
    \zeta(\ell) = 4 \pi f_V \int_{E_{\rm min}}^{E_{\rm max}} j(E) \sigma(E) dE
\end{equation}
where $\sigma(E)$ is the total ionization cross section. We use the empirical fit from \citet{rudd1985}. 

\section{Results and Discussion}\label{sec:results}

\begin{figure*}
    \centering
    \includegraphics[width=\textwidth]{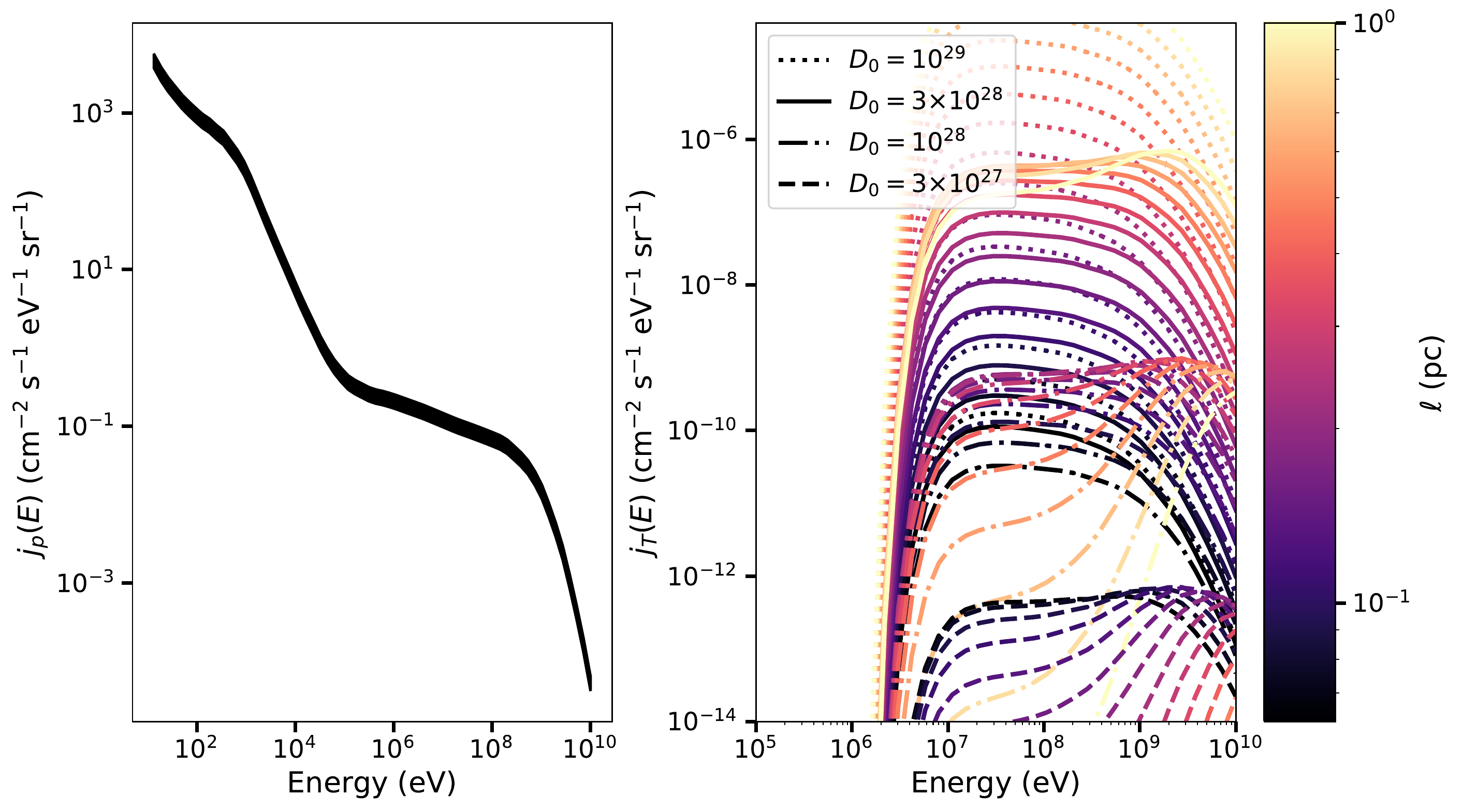}
    \caption{\label{fig:flx} Left: Steady state solution for the flux, $j_p(E)$ as a function of energy for different values of $\ell$. Right: Transported cosmic ray flux, $j_T(E)$, as a function of energy for different values of $D_0$ and $\ell$. Note the different scales for the x- and y-axis of the subplots. Without including the impact of energy-losses and diffusion throughout the cloud, the resulting cosmic-ray flux would produce nonphysically high ionization rates.}
\end{figure*}

For the following results, our canonical cloud is virialized, $\alpha_V = 1$, and turbulence is injected at a scale of 1 pc with a turbulent linewidth of 1 km s$^{-1}$. This results in the cloud primarily being sub-Alfv\'{e}nic \citep[see e.g.][]{crutcher2019}. Figure \ref{fig:flx} shows the steady-state flux, $j_p(E)$, and the transported flux, $j_T(E)$, as a function of scale $\ell$ and CR energy, E. Both $Q(E)$ and $j_p(E)$ are weakly dependent on the length scale, $\ell$. The CRIR associated with $j_p(E)$ are of order $\zeta \approx 7\times 10^{-10} - 10^{-9}$ s$^{-1}$ (not shown in the figure). These CRIRs are too high to be physical, and highlight the necessity of treating the diffusion of the CRs throughout the rest of the cloud structures.

The final CR spectrum, $j_T(E)$, shows significant variation with both $D_0$ and $\ell$. The flux at low energies is dramatically decreased due to energy losses from ionizations, Coulomb interactions and pion production while at high energies the flux is suppressed by diffusion. Our model predicts reconnection will seed the cloud with protons of energies $E \approx 10^{6} - 10^{10}$ eV. The resulting spectrum is greatly sensitive to the diffusion coefficient, $D_0$. For diffusion coefficients $D_0 = 10^{29}$ and $3\times10^{28}$ cm$^2$ s$^{-1}$, the resulting spectrum is relatively flat. However, for lower values of $D_0$, the spectrum is only flat for reconnection driven by the smallest scales of turbulence. We find a significant change in behavior between high $D_0$ and low $D_0$ values. For $D_0 = 3\times 10^{28}$ and $10^{29}$ cm$^2$ s$^{-1}$, we find the flux typically increases with the turbulence driving length scale, $\ell$. For $D_0 = 10^{28}$ and $3\times 10^{27}$ cm$^2$ s$^{-1}$ the flux decreases with increasing length scales. Conversely, if the particles travel ballistically, the transport length $\ell_T \approx R$ and the CRIR increases dramatically to $\zeta \approx 10^{-14}$ s$^{-1}$ for the fiducial model. This CRIR is far outside the observed range within the Milky Way, except in sightlines towards the galactic center \citep{indriolo2015}. Therefore, in the framework of CR acceleration by magnetic reconnection, low energy CRs must travel diffusely or the acceleration efficiency must be $f_{\rm CR} \ll 0.01$ to be consistent with observations.

\begin{figure}
    \centering
    \plotone{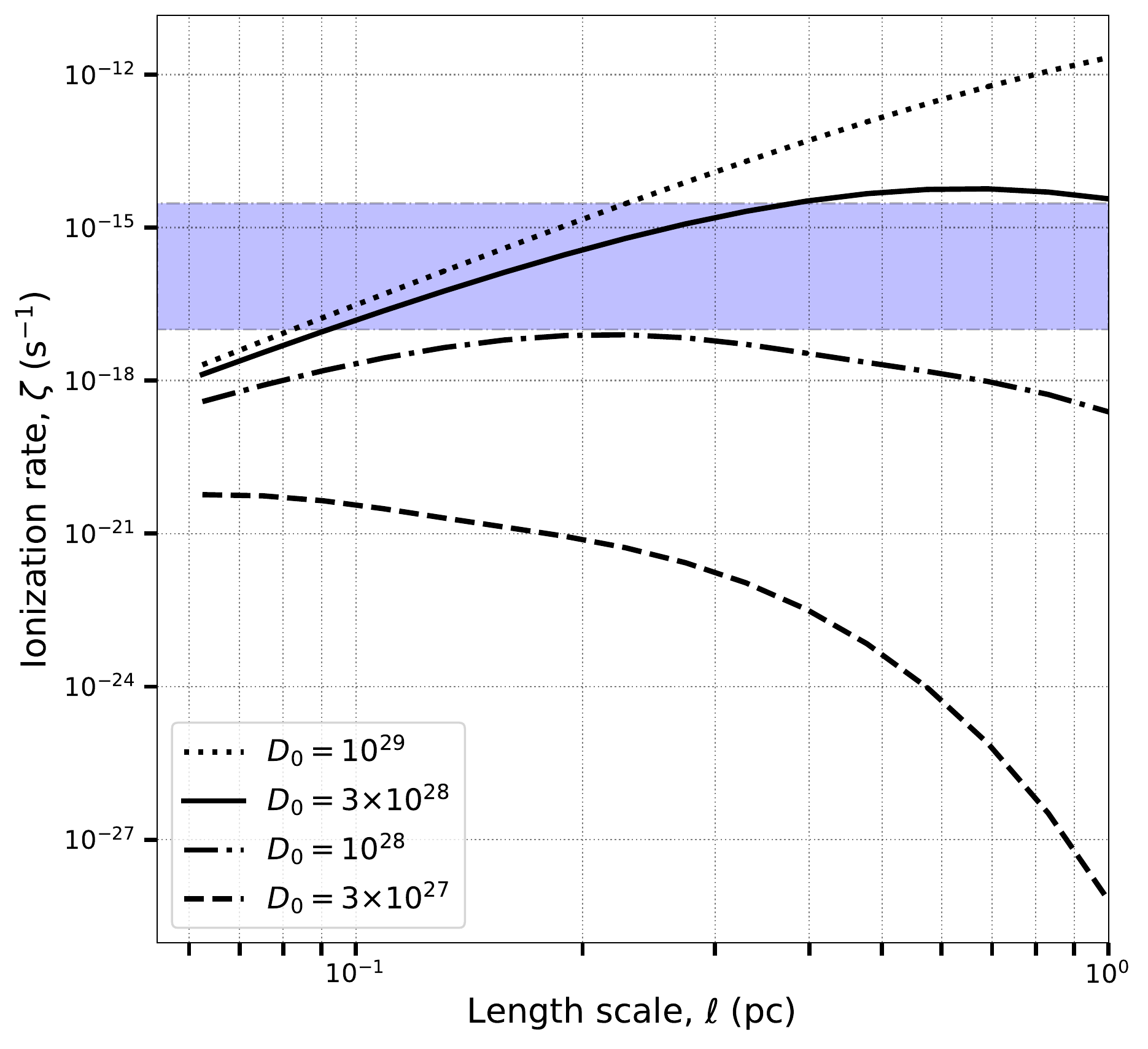}
    \caption{\label{fig:mainres} Left: Cosmic ray ionization rate, $\zeta(\ell)$, due to turbulent reconnection from turbulence at length scale, $\ell$, accounting for diffusion and energy losses.}
\end{figure}

Figure \ref{fig:mainres} shows the resulting CRIR, $\zeta(\ell)$ as a function of length scale, $\ell$, for different values of $D_0$. We highlight in the blue box the range of inferred values of $\zeta$ for molecular clouds in the Milky Way. We find that for values of $D_0 > 10^{28}$ cm$^2$ s$^{-1}$, our model is able to produce (or over-produce) the CRIR inferred in molecular clouds. For low values of $D_0$, the CRIR is below even the effective minimum ionization rate $\approx 10^{-19}$ in molecular clouds due to radioactive nuclei decay \citep{adams2014}. 

\begin{deluxetable}{c|CCCC}
\tablecaption{Power-Spectrum Averaged Cloud Ionisation Rates}
\tablenum{1}
\tablehead{\colhead{$D_0 = $} & \colhead{$3\times 10^{27}$} & \colhead{$1\times 10^{28}$} & \colhead{$3\times 10^{28}$} & \colhead{$1\times 10^{29}$}}
\startdata
Fiducial & \downarrow & 9.1(-19)  & 7.5(-17) & 2.5(-15) \\
Less Bound & \downarrow & 2.8(-18) & 2.3(-16) & 8.0(-15) \\
Strong Turb. & 2.9(-18) & 1.1(-15) & 6.2(-14) & 6.7(-13) \\
Kolmogorov Turb. & \downarrow & 1.1(-17) & 1.7(-16) & 1.6(-15) \\
\enddata
\tablecomments{\label{tab:crirks} The power-spectrum averaged cloud cosmic-ray ionization rates ($s^{-1}$), $\zeta_W$, (Eq. \ref{eq:sigw}) for different values of $D_0$. The value in the parenthesis indicates the power. The fiducial model uses the parameters $\alpha_V = 1$, $\beta = 0.5$, $\sigma_0 = 1$ km s$^{-1}$ and $L = 1$ pc. The rest of the rows delineate models with a specific parameter variation: ``Less Bound'' corresponds to $\alpha_V = 2$, ``Strong Turbulence'' to $\sigma_V = 2.5$ km s$^{-1}$ and ``Kolmogorov Turbulence'' to $\beta = 0.33$. The $\downarrow$ represents ionization rates below the radionuclide ionization rate \citep{adams2014}, $\zeta_{\rm RN} \approx 10^{-19}$.}
\end{deluxetable}

Table \ref{tab:crirks} shows the power-spectrum averaged CRIR, defined as
\begin{equation}\label{eq:sigw}
    \zeta_W = \frac{\int_{2\pi/L}^{k_{\rm max}} W(k) \zeta(\ell) k^2 dk}{\int_{2\pi/L}^{k_{\rm max}} W(k) k^2 dk}
\end{equation}
where $k = 2\pi/\ell$ and $W(k)$ is the isotropic kinetic energy turbulence spectrum
\begin{equation}
    W(k) = \frac{\rho \sigma(k)^2}{2k}
\end{equation}
and $\sigma(k) = \sigma_0 \left (\frac{k_0}{k} \right )^\beta$. We find that for all cases, increasing $D_0$ (and thus allowing CRs to propagate more easily throughout the cloud) systematically increases the CRIR. For the ``less bound'' clouds, the ionization rate increases due to the decreased average density, and hence CRs lose less energy through the cloud. Similarly, for both the ``strong turbulence'' model and ``Kolmogorov turbulence'' model, for which the turbulence strength is not increased, the produced CRIR is increased due to the enhanced turbulent power throughout the driving scales.

Our ``strong turbulence'' model represents regions of significant driving, such as in regions of enhanced star formation feedback \citep[e.g.][]{offner2018} (e.g., nearby protostar jets, high-mass stars and supernovae) or in the Galactic Center \citep{kauffmann2017}. Due to the strength of the turbulence, there is a significantly enhanced produced CRIR, far exceeding that observed in Solar neighborhood clouds. However, CRIRs on the order of 10$^{-14}$ s$^{-1}$ are observed through H$_3^+$ absorption towards the Galactic Center \citep{indriolo2015}. 

Most of the clouds in the Milky Way are not entirely virialized, and exhibit virial parameters greater than 1 \citep{heyer2015}. Therefore, our model predicts that within these clouds, reconnection within the MHD turbulence produces enough MeV – GeV protons to sustain CRIRs, $\zeta > 10^{-16}$ s$^{-1}$. 

This mechanism directly correlates the CRIR and the properties of the magneto-hydrodynamic turbulence within molecular clouds, along with the transport physics of low-energy CRs. Therefore, it may be possible to verify this mechanism with co-spatial observations of the ionization rate in dense gas, the magnetic field strength and the turbulence properties through observations of molecular ions and dust polarization maps. However, inferring the CRIR from such observations will rely on understanding the diffusion coefficient. Conversely, if the CRIR is dominated in dense gas by our proposed mechanism, it may be possible to infer the properties of the magneto-hydrodynamic turbulence from the CRIR through backwards modelling.

It is worth discussing the great uncertainty in the diffusion coefficient. Within the Milky Way, cosmic-ray transport studies and observations have indicated an average diffusion coefficient between $D_0=10^{28}$ -- $3\times10^{28}$ cm$^2$ s$^{-1}$ \citep{evoli2020}. However, regarding the dense gas, studies have shown a spread over several orders of magnitude, from $D_0 =10^{27}$ -- $10^{30}$ cm$^2$ s$^{-1}$ \citep{dogiel2015,owen2021}. As such, it is even more paramount to understand what is constraining the CR transport within molecular gas, and how the local environment changes the diffusion coefficient. Therefore, a widespread is observed, although dwarf galaxies appear to necessitate a higher ionization rate. If this is the case, our model predicts that dwarf galaxies would have CRIRs significantly greater than Milky Way-type galaxies.

We have proposed a novel mechanism for the production of low-energy CRs in molecular clouds through Fermi acceleration in regions undergoing turbulence reconnection: CRAFT. Since the MHD turbulence cascades across a wide range of scales, and since both the turbulence and the reconnection are volume-filling processes, we expect this will produce an approximately isotropic and homogenous floor to the CRIR. Historically, there has been a contradiction between the constant CRIRs used in astrochemical models \citep{rollig2007} and theoretical calculations, which have ubiquitously shown that energy losses would produce steep gradients with a low CRIR towards the cloud’s center. Furthermore, observations indicate the ionization rate in dense gas is not significantly lower than more diffuse regions. Our results would instead show that a properly chosen constant CRIR may be actually appropriate when modeling the dense gas in molecular clouds.

\acknowledgments
We thank the anonymous referee for their useful comments improving this work. This work was funded by theww ERC starting grant No. 679852 ‘RADFEEDBACK’. AL acknowledges the support by NASA TCAN AAG1967 and NSF AST 1816234. B.A.L.G would also like to thank his canine office mate, Mojo Gaches, for providing constant support during this work during the Coronavirus pandemic, although due to constantly sleeping on the job is not a co-author.

\bibliographystyle{aasjournal}
\bibliography{lib}

\end{document}